\documentclass{JAC2000}
\usepackage{graphicx}
\newcommand{\ud}{\mathrm{d}}
\begin{document}
\title{MULTIRESOLUTION REPRESENTATIONS FOR SOLUTIONS OF VLASOV-MAXWELL-POISSON
EQUATIONS}
\author{A. Fedorova,  M. Zeitlin, 
IPME, RAS, V.O. Bolshoj pr., 61, 199178, St.~Petersburg, Russia 
\thanks{e-mail: zeitlin@math.ipme.ru}\thanks{ http://www.ipme.ru/zeitlin.html;
http://www.ipme.nw.ru/zeitlin.html}
}

\maketitle

\begin{abstract}
We present the applications of variational--wavelet approach for computing 
multiresolution/multiscale representation for solution of some
approximations of Vlasov-Maxwell-Poisson equations.
\end{abstract}

\section{INTRODUCTION}
In this paper we consider the applications of a new nu\-me\-ri\-cal\--analytical 
technique which is based on the methods of local nonlinear harmonic
analysis or wavelet analysis to the nonlinear beam/accelerator physics
problems described by some forms of Vlasov-Maxwell-Poisson equations.
Such approach may be useful in all models in which  it is 
possible and reasonable to reduce all complicated problems related with 
statistical distributions to the problems described 
by systems of nonlinear ordinary/partial differential 
equations with or without some (functional)constraints.
Wavelet analysis is a relatively novel set of mathematical
methods, which gives us the possibility to work with well-localized bases in
functional spaces and gives for the general type of operators (differential,
integral, pseudodifferential) in such bases the maximum sparse forms. 
Our approach in this paper is based on the 
variational-wavelet 
approach from [1]-[10],
which allows us to consider polynomial and rational type of 
nonlinearities.
The solution has the following multiscale/multiresolution decomposition via 
nonlinear high-localized eigenmodes 
\begin{eqnarray}\label{eq:z}
u(t,x)&=&\sum_{k\in Z^2}U^k(x)V^k(t),\\
V^k(t)&=&V_N^{k,slow}(t)+\sum_{i\geq N}V^k_i(\omega^1_it), \quad \omega^1_i\sim 2^i \nonumber\\
U^k(x)&=&U_M^{k,slow}(x)+\sum_{j\geq M}U^k_j(\omega^2_jx), \quad \omega^2_j\sim 2^j \nonumber
\end{eqnarray}
which corresponds to the full multiresolution expansion in all time/space 
scales.

Formula (\ref{eq:z}) gives us expansion into the slow part $u_{N,M}^{slow}$
and fast oscillating parts for arbitrary N, M.  So, we may move
from coarse scales of resolution to the 
finest one for obtaining more detailed information about our dynamical process.
The first term in the RHS of formulae (1) corresponds on the global level
of function space decomposition to  resolution space and the second one
to detail space. In this way we give contribution to our full solution
from each scale of resolution or each time/space scale or from each nonlinear eigenmode
(Fig.1). 
The same is correct for the contribution to power spectral density
(energy spectrum): we can take into account contributions from each
level/scale of resolution.
Starting  in part 2 from Vlasov-Maxwell-Poisson equations
we consider in part 3 the approach based on
variational-wavelet formulation in the bases of compactly
supported wavelets or nonlinear eigenmodes. 
\begin{figure}[htb]
\centering
\includegraphics*[width=50mm]{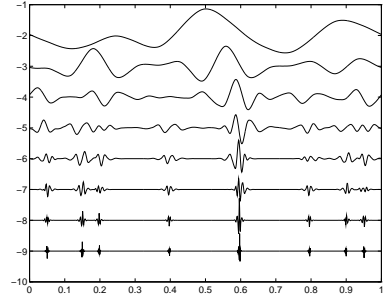}
\caption{Multiscale/eigenmode decomposition.}
\end{figure}

\section{Vlasov-Maxwell-Poisson Equations}

Analysis based on the non-linear Vlasov-Maxwell-Poisson equations leds to more
clear understanding of the collecti\-ve effects and nonlinear beam dynamics
of high intensity beam propagation in periodic-focusing and uniform-focusing
transport systems.
We consider the following form of equations ([11] for setup and designation): 
\begin{eqnarray}
&&\Big\{\frac{\partial}{\partial s}+p_x\frac{\partial}{\partial x}+
             p_y\frac{\partial}{\partial y}-
\Big[k_x(s)x+\frac{\partial\psi}{\partial x}\Big]\frac{\partial}{\partial p_x}-\nonumber\\
&& \Big[k_y(s)y+\frac{\partial\psi}{\partial y}\Big]\frac{\partial}{\partial p_y}
  \Big\} f_b(x,y,p_x,p_y,s)=0, \\
&&\Big(\frac{\partial^2}{\partial x^2}+\frac{\partial^2}{\partial y^2}\Big)\psi=
-\frac{2\pi K_b}{N_b}\int \ud p_x \ud p_y f_b,\\
&&\int\ud x\ud y\ud p_x\ud p_y f_b=N_b
\end{eqnarray}
The corresponding Hamiltonian for transverse sing\-le\--par\-ticle motion is given by 
\begin{eqnarray}
&& H(x,y,p_x,p_y,s)=\frac{1}{2}(p_x^2+p_y^2) 
                   +\frac{1}{2}[k_x(s)x^2 \\
 &&+k_y(s)y^2]+
    H_1(x,y,p_x,p_y,s)+\psi(x,y,s), \nonumber
\end{eqnarray}
where $H_1$ is nonlinear (polynomial/rational) part of the full Hamiltonian.
In case of Vlasov-Maxwell-Poisson system we may transform (2) into invariant form
 \begin{eqnarray}
\frac{\partial f_b}{\partial s}+[f,H]=0.
\end{eqnarray}

\section{Variational Multiscale Representation}

The first main part of our consideration is some variational approach, 
which reduces initial problem to the problem of
solution of functional equations at the first stage and some
algebraical problems at the second stage.
Multiresolution expansion is the second main part of our construction.
Because affine
group of translation and dilations is inside the approach, this
method resembles the action of a microscope. We have contribution to
final result from each scale of resolution from the whole
infinite scale of increasing closed subspaces $V_j$:
$\quad
...V_{-2}\subset V_{-1}\subset V_0\subset V_{1}\subset V_{2}\subset ...
$.
The solution is parameterized by solutions of two reduced algebraical
problems, one is nonlinear and the second are some linear
problems, which are obtained by the method of Connection
Coefficients (CC)[12].
We use compactly supported wavelet basis.
Let our  wavelet
expansion be
\begin{eqnarray}
f(x)=\sum\limits_{\ell\in{\bf Z}}c_\ell\varphi_\ell(x)+
\sum\limits_{j=0}^\infty\sum\limits_{k\in{\bf
Z}}c_{jk}\psi_{jk}(x)
\end{eqnarray}
If $c_{jk}=0$ for $j\geq J$, then $f(x)$ has an alternative
expansion in terms of dilated scaling functions only
$
f(x)=\sum\limits_{\ell\in {\bf Z}}c_{J\ell}\varphi_{J\ell}(x)
$.
This is a finite wavelet expansion, it can be written solely in
terms of translated scaling functions.
To solve our second associated linear problem we need to
evaluate derivatives of $f(x)$ in terms of $\varphi(x)$.
Let be $
\varphi^n_\ell=\ud^n\varphi_\ell(x)/\ud x^n
$.
We consider computation of the wavelet - Galerkin integrals.
Let $f^d(x)$ be d-derivative of function
 $f(x)$, then we have
$
f^d(x)=\sum_\ell c_l\varphi_\ell^d(x)
$,
and values $\varphi_\ell^d(x)$ can be expanded in terms of
$\varphi(x)$
\begin{eqnarray}
\varphi_\ell^d(x)&=&\sum\limits_m\lambda_m\varphi_m(x),\\
\lambda_m&=&\int\limits_{-\infty}^{\infty}\varphi_\ell^d(x)\varphi_m(x)\ud x,\nonumber
 \end{eqnarray}
where $\lambda_m$ are wavelet-Galerkin integrals.
The coefficients $\lambda_m$  are 2-term connection
coefficients. In general we need to find $(d_i\geq 0)$
\begin{eqnarray}
\Lambda^{d_1 d_2 ...d_n}_{\ell_1 \ell_2 ...\ell_n}=
 \int\limits_{-\infty}^{\infty}\prod\varphi^{d_i}_{\ell_i}(x)dx
\end{eqnarray}
For quadratic nonlinearities we need to evaluate two and three
connection coefficients
\begin{eqnarray}
&&\Lambda_\ell^{d_1
d_2}=\int^\infty_{-\infty}\varphi^{d_1}(x)\varphi_\ell^{d_2}(x)dx,
\\
&&\Lambda^{d_1 d_2
d_3}=\int\limits_{-\infty}^\infty\varphi^{d_1}(x)\varphi_
\ell^{d_2}(x)\varphi^{d_3}_m(x)dx \nonumber
\end{eqnarray}
According to CC method [12] we use the next construction. When $N$  in
scaling equation is a finite even positive integer the function
$\varphi(x)$  has compact support contained in $[0,N-1]$.
For a fixed triple $(d_1,d_2,d_3)$ only some  $\Lambda_{\ell
 m}^{d_1 d_2 d_3}$ are nonzero: $2-N\leq \ell\leq N-2,\quad
2-N\leq m\leq N-2,\quad |\ell-m|\leq N-2$. There are
$M=3N^2-9N+7$ such pairs $(\ell,m)$. Let $\Lambda^{d_1 d_2 d_3}$
be an M-vector, whose components are numbers $\Lambda^{d_1 d_2
d_3}_{\ell m}$. Then we have the first reduced algebraical system
: $\Lambda$
satisfy the system of equations $(d=d_1+d_2+d_3)$
\begin{eqnarray}
&&A\Lambda^{d_1 d_2 d_3}=2^{1-d}\Lambda^{d_1 d_2 d_3},
\\
&&A_{\ell,m;q,r}=\sum\limits_p a_p a_{q-2\ell+p}a_{r-2m+p}\nonumber
\end{eqnarray}
By moment equations we have created a system of $M+d+1$
equations in $M$ unknowns. It has rank $M$ and we can obtain
unique solution by combination of LU decomposition and QR
algorithm.
The second  reduced algebraical system gives us the 2-term connection
coefficients ($d=d_1+d_2$):
\begin{eqnarray}
A\Lambda^{d_1 d_2}=2^{1-d}\Lambda^{d_1 d_2},\quad 
A_{\ell,q}=\sum\limits_p a_p a_{q-2\ell+p}
\end{eqnarray}
For nonquadratic case we have analogously additional linear problems for
objects (9).
Solving these linear problems we obtain the coefficients of reduced nonlinear
algebraical system and after that we obtain the coefficients of wavelet
expansion (7). As a result we obtained the explicit time solution  of our
problem in the base of compactly supported wavelets. Also
in our case we need to consider
the extension of this approach to the case of
any type of variable coefficients (periodic, regular or singular).
We can produce such approach if we add in our construction additional
refinement equation, which encoded all information about variable
coefficients [13].
So, we need to compute only additional
integrals of
the form
\begin{equation}\label{eq:var1}
\int_Db_{ij}(t)(\varphi_1)^{d_1}(2^m t-k_1)(\varphi_2)^{d_2}
(2^m t-k_2)\ud x,
\end{equation}
where  $b_{ij}(t)$ are arbitrary functions of time and trial
functions $\varphi_1,\varphi_2$ satisfy the refinement equations:
\begin{equation}
\varphi_i(t)=\sum_{k\in{\bf Z}}a_{ik}\varphi_i(2t-k)
\end{equation}
If we consider all computations in the class of compactly supported wavelets
then only a finite number of coefficients do not vanish. To approximate
the non-constant coefficients, we need choose a different refinable function
$\varphi_3$ along with some local approximation scheme
\begin{equation}
(B_\ell f)(x):=\sum_{\alpha\in{\bf Z}}F_{\ell,k}(f)\varphi_3(2^\ell t-k),
\end{equation}
where $F_{\ell,k}$ are suitable functionals supported in a small neighborhood
of $2^{-\ell}k$ and then replace $b_{ij}$ in (\ref{eq:var1}) by
$B_\ell b_{ij}(t)$. In particular case one can take a characteristic function
and can thus approximate non-smooth coefficients locally. To guarantee
sufficient accuracy of the resulting approximation to (\ref{eq:var1})
it is important to have the flexibility of choosing $\varphi_3$ different
from $\varphi_1, \varphi_2$. In the case when D is some domain, we
can write
\begin{equation}
b_{ij}(t)\mid_D=\sum_{0\leq k\leq 2^\ell}b_{ij}(t)\chi_D(2^\ell t-k),
\end{equation}
where $\chi_D$ is characteristic function of D. So, if we take
$\varphi_4=\chi_D$, which is again a refinable function, then the problem of
computation of (\ref{eq:var1}) is reduced to the problem of calculation of
integral
\begin{eqnarray}
&&H(k_1,k_2,k_3,k_4)=H(k)=
\int_{{\bf R}^s}\varphi_4(2^j t-k_1)\cdot \nonumber\\
&&\varphi_3(2^\ell t-k_2)
\varphi_1^{d_1}(2^r t-k_3)
\varphi_2^{d_2}(2^st-k_4)\ud x
\end{eqnarray}
The key point is that these integrals also satisfy some sort of refinement
equation [13]:
\begin{equation}
2^{-|\mu|}H(k)=\sum_{\ell\in{\bf Z}}b_{2k-\ell}H(\ell),\qquad \mu=d_1+d_2.
\end{equation}
This equation can be interpreted as the problem of computing an eigenvector.
Thus, the problem of extension of the case of
variable coefficients are reduced to the same standard 
algebraical problem as in
case of constant coefficients. So, the general scheme is the same one and we
have only one more additional
linear algebraic problem by which we can parameterize the
solutions of corresponding problem in the same way.

So, we use wavelet bases with their good space/time      
localization properties to explore the dynamics of coherent structures in      
spa\-ti\-al\-ly\--ex\-te\-nd\-ed stochastic systems.
After some ansatzes, reductions and constructions we give for (2)-(6)
the following representation for solutions
\begin{equation}                                             
u(z,s)=\sum_{k}\sum_{\ell} U_{\ell}^k(z)V_\ell^k(s)=          
\sum U_\ell^kV_\ell^k,                                             
\end{equation}                                                           
where $V_\ell^k(s)$,                           
 $U_\ell^k(z)$ are both wavelets or nonlinear high-localized eigenmodes
and $z=(x,y)$.                        

Resulting multiresolution/multiscale representation for solutions
of (2)-(6) in the high-localized bases
is demonstrated on Fig.2.
\begin{figure}
\centering
\includegraphics*[width=60mm]{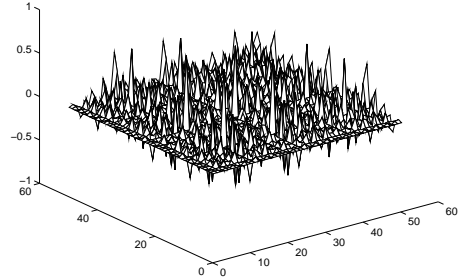}
\caption{The solution via multiscales.}
\end{figure}

We would like to thank Prof.
J.B. Rosenzweig and Mrs. Melinda Laraneta (UCLA) and
Prof. M.Regler (IHEP, Vienna) for
nice hospitality, help and support during UCLA ICFA Workshop and EPAC00.

 \end{document}